%% file: whitbeck_pmc10.tex
\documentclass[preprint]{elsarticle}

\usepackage{hyperref}
\usepackage[caption=false,font=footnotesize]{subfig}
\usepackage{array}
\usepackage{float}
\usepackage{url}
\usepackage{microtype}

\begin{document}

\title{From Encounters to Plausible Mobility\tnoteref{mobiopp}}

\tnotetext[mobiopp]{This article is an expanded version of work presented at the 2\textsuperscript{nd} ACM/SIGMOBILE Workshop on Mobile Opportunistic Networking (MobiOpp 2010)~\cite{whitbeck-mobiopp10}. This paper presents all-new results based on an improved mobility inference heuristic. The \emph{plausible mobility} approach is much more thoroughly evaluated using both synthetic and experimental mobility traces.}

\fntext[acknowledgments]{Work partially done while this author was visiting the College of Computing at GeorgiaTech. This work has been partially supported by the ANR project Crowd under contract ANR-08-VERS-006 and NSF grant CNS0831714.}

\author[thales,lip6]{John~Whitbeck\fnref{acknowledgments}}
\ead{john.whitbeck@lip6.fr}

\author[lip6]{Marcelo~Dias~de~Amorim}
\ead{marcelo.amorim@lip6.fr}

\author[thales]{Vania~Conan}
\ead{vania.conan@fr.thalesgroup.com}

\author[gatech]{Mostafa~Ammar}
\ead{ammar@cc.gatech.edu}

\author[gatech]{Ellen~Zegura}
\ead{ewz@cc.gatech.edu}

\address[thales]{Thal\`es Communications\\
  160 bd de Valmy~-- 92704 Colombes Cedex~-- France}
\address[lip6]{LIP6/CNRS~-- UPMC Sorbonne Universit\'es \\
  104 avenue du Pr\'esident Kennedy~-- 75016 Paris~-- France}
\address[gatech]{College of Computing~-- Georgia Institute of Technology \\
  Atlanta, Georgia~-- 30332-0280~-- USA}

\begin{abstract}
Inferring plausible node mobility based only on information from wireless contact traces is a difficult problem. Working with mobility information allows richer protocol simulations, particularly in dense networks, but requires complex set-ups to measure. On the other hand, contact information is easier to measure but only allows for simplistic simulation models. In a contact trace a lot of node movement information is irretrievably lost so the original positions and velocities are in general out of reach. In this paper, we propose a fast heuristic algorithm, inspired by dynamic force-based graph drawing, capable of inferring a plausible movement from any contact trace, and evaluate it on both synthetic and real-life contact traces. Our results reveal that (i) the quality of the inferred mobility is directly linked to the precision of the measured contact trace, and (ii) the simple addition of appropriate anticipation forces between nodes leads to an accurate inferred mobility.
\end{abstract}


\maketitle

\input{intro}

\input{related}
\input{inferring}

\input{formalization}

\input{heuristic}

\input{evaluation_synthetic}
\input{evaluation_reallife}
\input{conclusion}

\bibliographystyle{elsarticle-num}
\bibliography{whitbeck_pmc10}

\end{document}

%% file: intro.tex
\section{Introduction}
\label{sec:intro}

In the disruption-tolerant network (DTN) paradigm, mobile communication devices undergo a sequence of connections and disconnections from other devices forming \textit{contact opportunities}~\cite{dtn_fall_sigcomm}. Despite the growing interest in exploiting these contact opportunities for disseminating information under conditions when more traditional approaches are either impractical or unfeasible, there have been few real-life DTN deployments~\cite{Zhang:2007,kiosknet}. Instead, most evaluations of new protocols and designs have been done through simulations based on either synthetic mobility models or real-life contact traces.

On the one hand, synthetic mobility models give full knowledge of the mobility and therefore allow for simulation of the specific features of radio channels (e.g., interferences and hidden stations) but do not accurately represent real-life mobility. On the other hand, contact traces are assumed to accurately represent real-life mobility but all geographical information is lost and simulators must make very simplistic assumptions on the communication channel (e.g., a node may only communicate with one of its current ``contacts'' at any given time~\cite{ONE}). A way out of this alternative could be the use of GPS measurements of human mobility~\cite{rhee:levy,piorkowski_hotplanet09}. Unfortunately, on top of not working indoors, these traces are typically so sparse that they fail to capture the contact opportunities between people. This may be due to a very small number of GPS devices~\cite{rhee:levy} or, a lack of contemporaneous paths, as in the Nokia Sports Tracker traces~\cite{nokia_sports_tracker}. Even if large scale GPS measurements, such as Google Latitude~\cite{google_latitude}, were able to achieve sufficient density, accurate propagation models and other technology-related characteristics would be required to identify the contact opportunities. Ideally, experiments should measure both contacts and positions, but this can be costly and sometimes unfeasible.

When only measuring the contact opportunities from an experiment with mobile devices, a lot of information is irretrievably lost. Consider a simple example with two nodes. When in contact, we can roughly locate them relatively to one another. However, when the time elapsed since the latest contact (inter-contact time) increases, the information regarding their relative distance decreases. After a while, it becomes difficult to say if they are still somewhat close or if they have gone in completely opposite directions. In a dense network, the higher the contact intensity, the more constrained our problem is. Although it is difficult to infer a mobility that is strongly correlated with the original mobility, we show in this paper that it is possible to propose a \textit{plausible mobility}, i.e., one that would have generated the same contact trace. This is examined in more detail in Section~\ref{sec:infering}. Whether the ultimate goal is to visualize, modify the original contact trace, or improve simulations, inferring the exact mobility is not required. All that is needed is a plausible inferred mobility that leads to realistic modifications or better predictions.

\textit{What if the information from the contact traces were sufficient to infer plausible node mobility?} If this were possible, there would be several immediate benefits. Firstly, being able to visualize node movement is in itself valuable, as it confers an intuitive understanding of the trace dynamics that can get lost in statistics. Secondly, using the inferred movement instead of simply contacts history allows for realistic \textit{transformations} of the contact trace, such as adding nodes, modifying the node density, or increasing communication range. Finally, this would allow for a much finer simulation of the radio channel, particularly for dense contact traces~\cite{tournoux08_rollernet}, while retaining the realism captured by the contact traces. This paper makes the following contributions:

\begin{itemize}

    \item We define and discuss the problem of inferring plausible node mobility from only their contact information. To the best of our knowledge, this is the first time such a problem is addressed.

    \item We propose a formal definition of the problem as a system of non-linear inequalities.

    \item We describe and evaluate, both on synthetic and real-life contact traces, a heuristic but practical and effective method of inferring the movement of the nodes. This method works offline and assumes full knowledge of past and future contacts.

\end{itemize}

In the next section, we position our paper in comparison with prior work. In Sections~\ref{sec:infering} and~\ref{sec:formal}, we formally define the problem of inferring mobility from contact traces and discuss its challenges. In Section~\ref{sec:heuristic} we propose a heuristic approach to efficiently solve our problem, which we then evaluate using both synthetic (Section~\ref{sec:evaluation_synthetic}) and real-life contact traces (Section~\ref{sec:evaluation_reallife}). Finally we conclude our work and discuss the path ahead in Section~\ref{sec:conclusion}.

%% file: related.tex
\section{Related Work}
\label{sec:related}

Delay/disruption-tolerant networks (DTN)~\cite{dtn_fall_sigcomm} arise when lack of end-to-end connectivity, rapidly changing topology, and/or potentially long communication delays render traditional mobile ad-hoc networks (MANET) approaches unfeasible~\cite{antonellis_classification}. Such networks encompass a vast spectrum of situations ranging from inter-planetary communications~\cite{dtn-interplanetary} to hop-by-hop data forwarding between portable devices to supplement an infrastructure for content dissemination~\cite{ioannidis09optimal}. In DTNs, node mobility can be exploited to increase the network capacity while compromising on delays by using a message \textit{store-and-forward} paradigm instead of the usual packet switching~\cite{GrossglauserTse2002}.

Opportunistic mobile networks are a class of DTNs in which no knowledge of the future mobility of nodes is assumed. For example, this is the case of a network formed by the direct contact opportunities of hand-held devices, such as smartphones. Contact opportunities between mobile devices could be used either to replace or assist a wireless infrastructure for the dissemination of a given content. For example, Ioannidis et al. studied how to combine content pushing from a source in the infrastructure with opportunistic forwarding among subscribers in a way that ensures perceived content freshness from the subscribers while keeping the load on the infrastructure as small as possible~\cite{ioannidis09optimal}.

\begin{table}
  \centering
  \caption{Some wireless contact traces.}
  \includegraphics{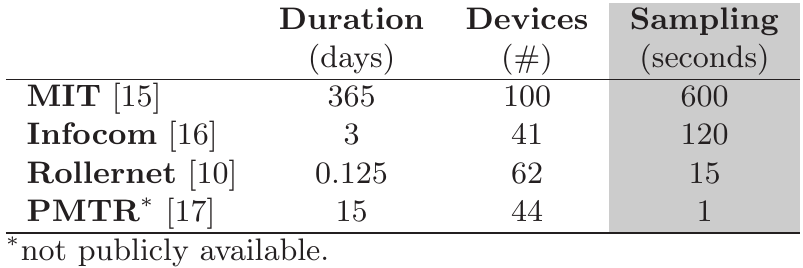}
  \label{table:traces}
\end{table}

A lot of effort has therefore gone into measuring human mobility, or at least the contact opportunities generated by human mobility. Direct measurements of human GPS coordinates have, to the best of our knowledge, only been performed by Rhee et al. in their work on human mobility models~\cite{rhee:levy}, and by commercial applications such as the Nokia Sports Tracker. Vehicular GPS measurements, from   taxis~\cite{comsnets09piorkowski} or buses~\cite{UMassDieselNet} are more common. These GPS measurements give accurate and fine-grained (10 and 1 second periods, respectively) information but unfortunately only work when outdoors, and, more importantly, the traces are too sparse to create any contact opportunities. 

To counter this problem, several strategies have been used. Piorkowski \textit{et al.} suppose that the movement of each node is stationary and ergodic, and that they collectively constitute a representative subset of the larger population~\cite{piorkowski_hotplanet09,comsnets09piorkowski}. In order to obtain a reasonable density, they shift the starting times of the measured mobility of various participants in order to make them all contemporaneous~\cite{piorkowski_hotplanet09}. Alternatively, they \emph{densify} the trace by adding many new nodes with mobility patterns that are statistically similar to measured ones~\cite{comsnets09piorkowski}. Furthermore, given a certain radio range, one could scale down a GPS trace in order to achieve a target node density.

Direct measurements of human contact opportunities overcome these limitations but forgo location information. In the Reality Mining experiment conducted at MIT, each participant had a special application running on her/his mobile phone that captured proximity information from 100 subjects over an academic year~\cite{mit}. The Haggle project used Intel iMotes to capture the contacts between participants of the Infocom 2005 conference~\cite{chaintreau_mobility}. Both of these experiments relied on periodic Bluetooth scans.

While Bluetooth has the advantage of being widely available, its scanning mechanism is too slow to effectively detect all contact opportunities. Indeed, the longer the sampling period (respectively 600 and 120 seconds for the MIT and Infocom traces), the more likely temporary link failures or short contacts are missed. The Rollernet experiment, which also uses iMotes to capture the contacts in a rollerblading tour, was able to bring the sampling period down to 15 seconds~\cite{tournoux08_rollernet}. For finer measurements, a different beaconing method must be used. For example, Gaito et al. designed a specific device, a Pocket Mobility Trace Recorder (PMTR), and were able to measure contact opportunities every second~\cite{PMTR}, but the traces are not yet publicly available. Table~\ref{table:traces} compares these different traces. As we will see in Section~\ref{sec:heuristic}, shorter sampling periods translate into more constraints on our mobility inference problem, which in turn lead to solutions that better match the original mobility.

Inferring node mobility based solely on contact information is an open problem that has not yet received much attention in opportunistic networks. Walker uses persistent homologies to characterize the topology of the surface on which a given contact trace was collected~\cite{walker_homology}. Using this technique, it is possible to identify if participants in a contact trace were moving on a ``ring'' for example. Yoneki et al. have developed a visualization tool for opportunistic contact traces but do not go try to infer a realistic mobility from it~\cite{yoneki_vizu}.

However, some similar questions have been addressed in other contexts. In wireless sensors networks, sensors can estimate their position relatively to a small number of \textit{anchor nodes} (typically equipped with a GPS receiver) using either a variety of distance measurement techniques based on received signal strength or differences in beacon timings or simply contact information (i.e., range-free)~\cite{wsn_localization}. However such methods often use computationally expensive techniques which do not scale to a mobile environment~\cite{doherty_lmi}. Baggio et al. propose a lightweight Monte-Carlo localization scheme for mobile scenarios but it requires a very high density of anchor nodes~\cite{baggio_mcl}.

%% file: inferring.tex
\section{Inferring mobility}
\label{sec:infering}

In our approach, we must rely solely on contact information and assume no low-level information on distances between nodes. Furthermore, unlike most wireless sensor networks, our nodes are all mobile and the network is sparse and disconnected. Finally, and perhaps most importantly, being free of decentralization or real-time requirements, our calculations take place offline and with full knowledge of past and future contacts.

\subsection{Problem definition}
\label{subsec:problem}

Let us consider a fixed number of mobile nodes. A contact trace is the list of contact events that occur between these nodes. Each event is recorded as a quadruplet consisting of the identity of both nodes, the instant when the contact is first established, and the instant when the contact goes down.

\paragraph{Noisy real-life traces}
In real-life traces, depending on the scanning period and the choice of radio technology (e.g., Bluetooth, ZigBee), a number of contact opportunities may be missed, shortened, split, or merged. For example, using Bluetooth, neighborhood scans typically take several seconds and may not detect all reachable devices due to the frequency-hopping nature of the protocol. By using longer sampling periods, it becomes difficult to detect short contacts; even worse, a sequence of short contacts will likely be considered as a single long contact. Furthermore, a Bluetooth device may not simultaneously scan and respond to a scan. Therefore, many contacts will be missed simply because of the nature of the underlying protocol. Other wireless technologies, such as the custom-made Pocket Mobility Trace Recorders~\cite{PMTR}, can overcome these limitations but still have to contend with the traditional wireless issues such as interferences or hidden terminals. For these reasons, real-life traces must be considered noisy.

\paragraph{``Perfect'' synthetic contact traces}
Contact traces can also be extracted from synthetic mobility models by simulating a beaconing protocol or using a simple proximity-based model (i.e., a contact exists when two nodes are in transmission range of each other). Traces obtained in this fashion can be considered \emph{perfect}, in the sense that we have full control over all parameters and all contact opportunities are recorded. We will use this approach to evaluate our heuristic algorithm proposed in Section~\ref{sec:heuristic}.

\paragraph{Additional topology information}
Some nodes (i.e. \emph{anchor} nodes) could have known positions, such as base stations in a 3G/WiFi network or GPS-equipped mobile nodes, which enables us to place other nodes relatively to them. On static configurations, estimating positions based solely on connectivity information has been well studied in wireless sensor networks ~\cite{wsn_localization}. When all nodes are mobile, relative positioning information may still be available. For example, in the Rollernet experiment~\cite{tournoux08_rollernet}, an iMote was given to a member of staff that remained at front the rollerblading tour, while another was given to someone who stayed at the back. All other nodes in the trace must therefore be placed between these tail and head nodes. Finally, we could suppose that only the initial positions of the nodes are known. For synthetic traces, this information is readily available.

\paragraph{Plausible mobility}
Since we cannot hope to recover the exact initial mobility from a pure contact trace, we define the concept of \emph{plausible mobility}. In order to be \emph{plausible}, the inferred movement must (i) realistically limit the speed of the nodes and (ii) possibly produce the original contact trace, i.e. nodes in contact \emph{must} be within transmission range of each other while nodes not in contact \emph{should} be beyond transmission range. In the end, our objective is to develop an algorithm that takes a contact trace and some additional information (e.g., fixed or relative positions) as input and generate a \emph{plausible} movement trace as output.

\subsection{Evaluation framework}
\label{subsec:framework}

\begin{figure}[t]
  \centering
  \includegraphics{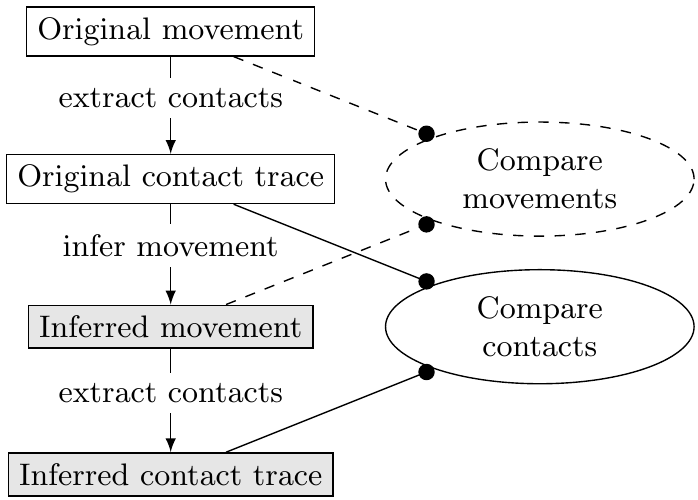}
  \caption{Evaluation framework.}
  \label{fig:evaluation}
\end{figure}

Fig.~\ref{fig:evaluation} outlines our evaluation framework. Basically, we consider two ways of evaluating a mobility inference method: one comparing original and inferred mobility, and the other comparing original and inferred contact traces. The former option is only available with synthetic traces while the latter is available for both synthetic and real-life traces. 

Indeed, When using synthetic mobility models, we initially have information of the nodes' mobility. From this, we can extract a contact trace, which we will use as input to our mobility inference algorithm. The inferred mobility can be compared to the original mobility, but can also be used to extract an inferred contact trace, which in turn can be compared to the original contact trace.

When using real-life contact traces, the original mobility is not available, and thus original and inferred mobility can no longer be compared. It is, however, still possible to compare the original and inferred contact traces.

%% file: formalization.tex
\section{Formalization}
\label{sec:formal}

In this section, we describe what would constitute an ideal inference of mobility. The constraints defined below will guide us in the choice of the parameters for the heuristic approach proposed in Section~\ref{sec:heuristic}. Since the input is a contact trace, complete knowledge of past, present, and future contacts is assumed (offline inference).

\subsection{Definitions}
\label{subsec:definitions}

\begin{table}
\centering
\caption{Notation.}
\includegraphics{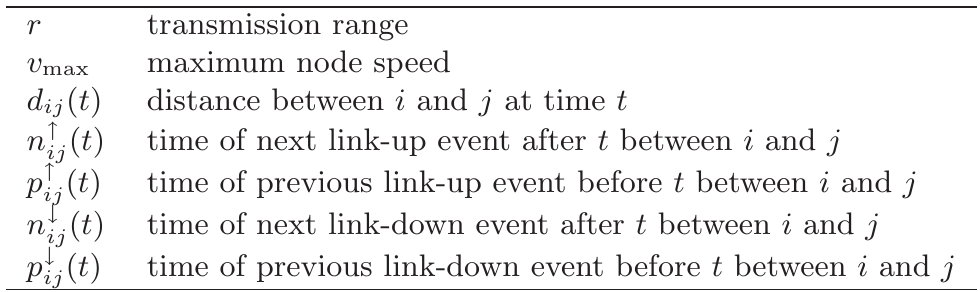}
\label{table:notations}
\end{table}

\begin{figure}
\centering
\includegraphics{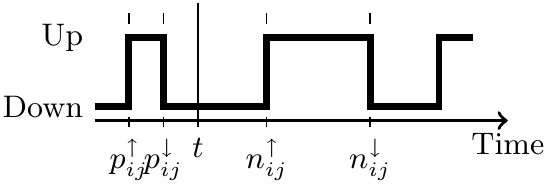} \quad \includegraphics{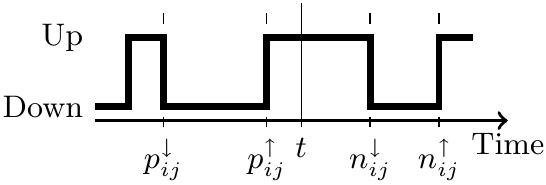}
\caption{Notation for link-up and link-down events.}
\label{fig:prev_next}
\end{figure}

Let $N$ be the number of mobile nodes in the contact trace. These nodes move on a 2D plane, have a maximum speed $v_{\max}$ and a transmission range $r$. Let $(x_i(t),y_i(t))$ be the coordinates of node $i \in \{1,\cdots,N\}$. For the pair of nodes $(i,j)$, let $d_{ij}(t)$ denote the Euclidian distance between $i$ and $j$ at time $t$.
At any given time $t$, let $n^{\uparrow}_{ij}(t)$ (resp. $p^{\uparrow}_{ij}(t)$) be the next (resp. previous) time the link between $i$ and $j$ comes up. Conversely, let $n^{\downarrow}_{ij}(t)$ (resp. $p^{\downarrow}_{ij}(t)$) be the next (resp. previous) time the link between $i$ and $j$ goes down. In the rest of this paper, their dependence on the current time $t$ is obvious and thus omitted for clarity. This notation is summed up in Table~\ref{table:notations}, and Fig.~\ref{fig:prev_next} illustrates the time notation.

At any time $t$ and any time interval $\Delta t$, the maximum node speed $v_{\max}$ imposes the following constraint on the positions of any node $i$:

\begin{equation}
\textrm{\textbf{Constraint~1: }} \sqrt{{(x_i(t+\Delta t) - x_i(t))}^2 + {(y_i(t+\Delta t)-y_i(t))}^2} \le v_{\max} \Delta t.
\label{eqn:spd_cnstr}
\end{equation}

This constraint imposes that, given a valid solution at time $t$ and a short time interval $\Delta t$, the next valid position at time $t+\Delta t$ should be very similar.

\subsection{Case 1: Synthetic contact traces}
\label{subsec:perfect}

\begin{figure}[t]
  \centering
  \includegraphics{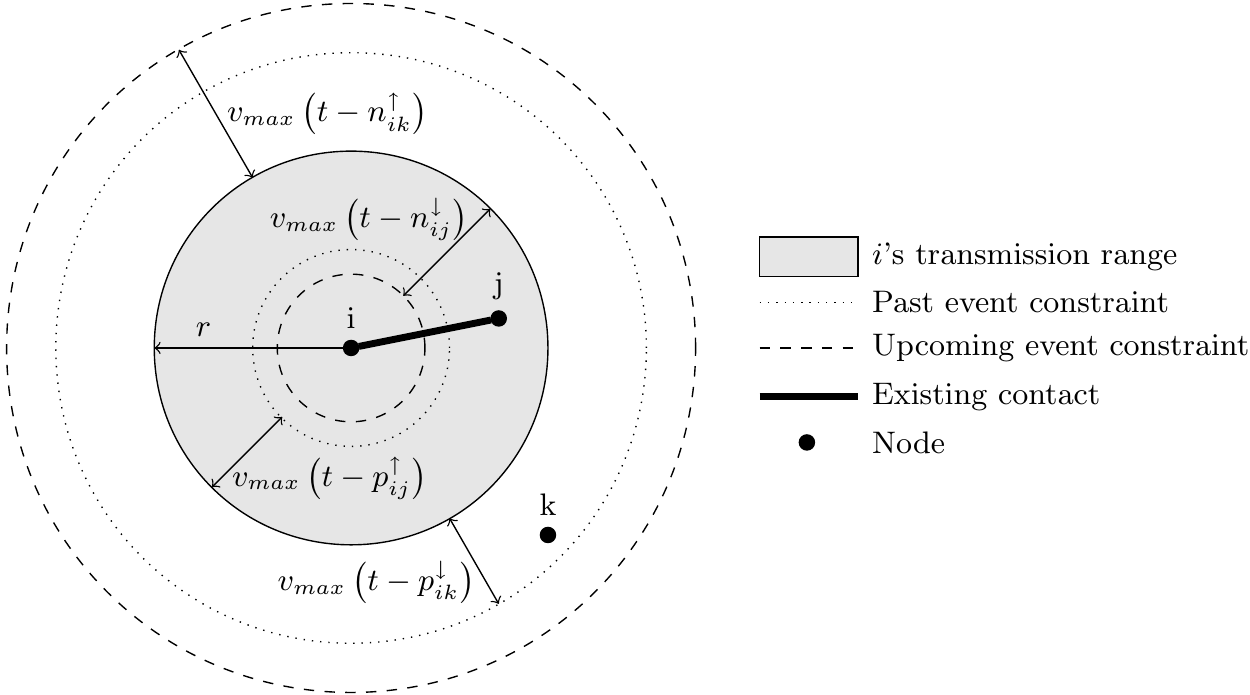}
  \caption{Constraints on the positions of nodes $j$ and $k$ relative to node $i$. The contact between nodes $i$ and $j$ came up at time $p^{\uparrow}_{ij}(t)$ and will go down at time $n^{\downarrow}_{ij}(t)$. The contact between nodes $i$ and $k$ came down at time $p^{\downarrow}_{ik}(t)$ and will come back up at time $n^{\uparrow}_{ik}(t)$. Parameters $r$ and $v_{\max}$ denote the transmission range and maximum speed, respectively. See Section~\ref{subsec:perfect}.}
  \label{fig:ranges}
\end{figure}

In a synthetic contact trace, a contact appears when the distance between two nodes is less than $r$ and breaks when it is greater than $r$. Since we know when current contacts are going to break and when new ones will appear, we can further constrain nodes' positions. Indeed, nodes must get closer to each other before the contact appears and move away from each other before it goes down. This is illustrated in Fig.~\ref{fig:ranges}, where the contact at time $t$ between nodes $i$ and $j$ appeared at time $p^\uparrow_{ij}$ and will go down at $n^\downarrow_{ij}$, and a contact between $i$ and $k$ went down at time $p^\downarrow_{ik}$ and will reappear at time $n^\uparrow_{ik}$.

Right after nodes $i$ and $j$ come into contact at time $p^\uparrow_{ij}$, their initial distance is $r$. Since their maximum speed is $v_{max}$, the distance between them cannot shrink faster than $2v_{\max}$ if both nodes are moving directly at each other. Conversely, as $t$ approaches the time $n^\downarrow_{ij}$, when the contact between $i$ and $j$ goes down, both nodes must be moving out of each other's transmission range. Relatively to $i$, $j$ must be able, given its maximal speed $v_{\max}$ to move out of transmission range at exactly $n^\downarrow_{ij}$. Therefore, if $i$ and $j$ are in contact at time $t$, the following constraint holds:

\begin{equation}
\textrm{\textbf{Constraint~2: }} r - 2 v_{\max} \textrm{min}\left\{t-p^\uparrow_{ij}, n^\downarrow_{ij}-t\right\} \le d_{ij} \le r.
\label{eqn:contact_cnstr}
\end{equation}

A similar analysis holds for the contact between $i$ and $k$. Their previous contact ended at time $p^\downarrow_{ik}$, and these two nodes cannot move apart faster than $2 v_{\max}$. As $t$ approaches the time $n^\uparrow_{ik}(t)$, when the contact between $i$ and $k$ will reappear, $k$ must come closer to $i$'s transmission range. Relatively to $i$, $k$ must be able to come within transmission range of $i$ at exactly $n^\uparrow_{ik}$. Therefore, while $i$ and $k$ are not in contact, the following constraint holds:

\begin{equation}
\textrm{\textbf{Constraint~3: }} r \le d_{ik}(t) \le r+2 v_{\max} \textrm{min}\left\{t-p^\downarrow_{ik}, n^\uparrow_{ik}-t\right\}.
\label{eqn:not_contact_cnstr}
\end{equation}

\subsection{Case 2: Real contact traces}
\label{subsec:real}

While we know that a movement satisfying constraints~(\ref{eqn:spd_cnstr}), (\ref{eqn:contact_cnstr}), and~(\ref{eqn:not_contact_cnstr}) exists for a synthetic contact trace (i.e., the original synthetic movement), this is less clear for a real-life trace. Indeed, as previously discussed, a real-life contact trace may be quite noisy and, in particular, miss many contacts. While this may seem like a simple relaxation of our constraints, it could in fact make the system unsolvable. Indeed, when considering real-life traces, the inclusive (i.e., \textit{in-contact}, Eq.~\ref{eqn:contact_cnstr}) and the exclusive (i.e., \textit{not-in-contact}, Eq.~\ref{eqn:not_contact_cnstr}) constraints no longer have the same importance. The inclusive constraint, based on the presence of a contact, can be trusted. However, the exclusive constraint, based on the absence of a contact, no longer strictly means that the distance between two nodes must be greater than the transmission range $r$. Indeed, one could imagine a node quickly passing by the other nodes, moving in and out their transmission ranges without triggering any contact detection. If we strictly enforce the exclusive constraint, such movements may no longer be possible.

%% file: heuristic.tex
\section{Heuristic solution}
\label{sec:heuristic}

In this section, we propose and evaluate a simple and efficient heuristic for inferring node mobility from their contact traces. Note that in order for it to have broad applicability, it should make as few assumptions as possible on the original mobility.

\subsection{Dynamic graph drawing}
\label{subsec:graph_drawing}

Our heuristic approach is inspired by works in \textit{dynamic graph animation}, even though its objective is quite different. Graph animation aims at (i) producing a sequence of readable and aesthetically pleasing representations of graphs and (ii) animating the transitions between successive graph layouts in a way that preserves the viewers' \textit{mental map} of the graph~\cite{onlineGD}. Sample applications include visualization of communication networks, social networks, and software library dependencies. Both goals are relevant to us. Not only do we wish to infer a sequence of positions for each node in the contact trace (i.e., a sequence of connectivity graph layouts), but real-life mobility intrinsically produces a sequence of connectivity graphs in which the transitions are easy to follow. However, while the function of dynamic graph animation is mostly aesthetic, our heuristic aims at meeting the constraints set out in Section~\ref{sec:formal}.

In the context of graph animation, \textit{online} means that the graph layout algorithm is continuously running while new nodes or links appear and disappear on the fly, whereas \textit{offline} means that each successive graph is laid out separately. The offline method makes it difficult to preserve the viewer's \textit{mental map} during transitions, particularly long ones, between successive graph layouts. The online method procures an illusion of continuous mobility and allows for easy control of nodes' speed but is, in itself, insufficient in our case. Indeed, when a contact occurs between two nodes, they may, at that time, be very far away from each other in the online graph animation. This leads to a link in the connectivity graph that will, at least temporarily, straddle several connected components, which cannot constitute a satisfactory inferred movement.

Of particular interest to us are the \textit{force-based} layout algorithms~\cite{Fruchterman91graphdrawing}, in which attractive and repulsive forces are applied to nodes according to the connectivity graph. As in a real physical system, the nodes then converge to a minimum stress (or energy) position. Force-based algorithms are particularly well suited to our problem because each pair of nodes that are in contact will tend to be geographically close to each other.

Our heuristic for inferring a \textit{plausible mobility} from a contact trace will consist of running an \textit{online force-based dynamic graph layout} algorithm, built from the forces and refinements described in the next two sections.

\subsection{Forces}
\label{subsec:forces}

\begin{figure}
  \centering
  \includegraphics{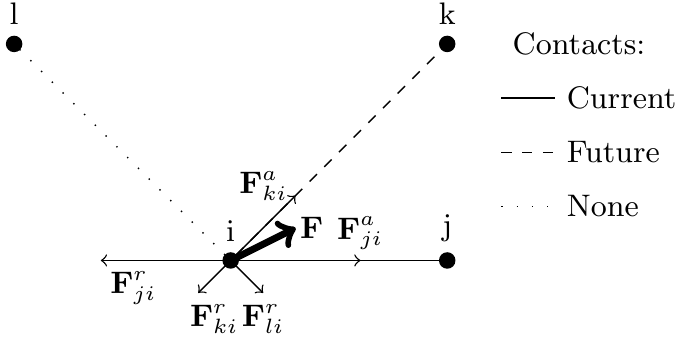}
  \caption{Forces applied to node $i$ from nearby nodes $j,k,l$.  $\mathbf{F}_{ji}^{a}$ is the attractive force of $j$ on $i$ and $\mathbf{F}_{ji}^{r}$ is a repulsive force. $\mathbf{F}$ is the resulting force applied to $i$, bringing $i$ closer to $k$ before the $\{i,k\}$ contact appears. Full explanation in Section~\ref{subsec:forces}.}
  \label{fig:forces}
\end{figure}

As in a physical system, each node has a position, a speed, and an acceleration. All nodes have the same mass. Between each pair of nodes is an \textit{attractive} and a \textit{repulsive} force. Optionally a \textit{drag} force may be added in certain circumstances. While the mathematical expression of these forces has its roots in force-based dynamic graph drawing, it has been modified to incorporate knowledge of past and future contacts. We first present these forces before discussing how to set their parameters in the next section. Fig.~\ref{fig:forces} shows how these forces add up. Hereafter, $\mathbf{u}_{ij}$ is the unitary vector directed from $i$ to $j$. Otherwise, the notation is the same as in Section~\ref{subsec:definitions} as summed up in Table~\ref{table:notations}.

\subsubsection{Attractive force}
Let $j$ be one of node $i$'s contacts (i.e., a neighbor in the connectivity graph). $j$ attracts $i$ with a spring-like force:

\begin{equation}
\mathbf{F}^{a}_{ji} =
\left\{
\begin{array}{ll}
K d_{ij} \mathbf{u}_{ik} & \textrm{if connected}, \\
K e^{-\frac{v_{\max} dt}{\tau}} d_{ij} \mathbf{u}_{ik}& \textrm{if not connected}, \\
\end{array}
\right .
\label{eqn:attractive}
\end{equation}

\noindent where $dt=\textrm{min}\{n^{\uparrow}_{ij}-t,t-p^{\downarrow}_{ij}\}$ and $K$ is a rigidity constant. When $i$ and $j$ are connected, this force acts as a classical ``spring'' force which contributes keeping to the two nodes within transmission range of each other (right part of the constraint(~\ref{eqn:contact_cnstr})). When not connected, we consider the time to the closest \textit{link-up} event, that is either the previous time $i$ and $j$ came into contact, or the next time they will come into contact. This ensures two things: (i) that after a contact goes down, there is still a lingering force preventing the two nodes from moving too quickly away from one another, and (ii) that, prior to a contact appearing, the two nodes will become gradually more attracted to each other and thus begin moving in each other's direction (right part of constraint(~\ref{eqn:not_contact_cnstr})). The intensity of this force is continuous over connection/disconnection events. The $\frac{v_{\max}}{\tau}$ fraction determines when upcoming or past events begin and cease to have a noticeable influence. This is discussed in more detail in Section~\ref{subsec:usage}.

\subsubsection{Repulsive force}
Each node $j \neq i$ pushes node $i$ back with a Coulomb-like force:

\begin{equation}
\mathbf{F}^{r}_{ji} =
\left\{
\begin{array}{ll}
- \frac{G}{\left(\epsilon_0+\frac{d_{ij}+v_{\max}dt}{r}\right)^\alpha} \mathbf{u}_{ik} & \textrm{if } d_{ij} < d_{\max}, \\
\mathbf{0} & \textrm{if } d_{ij} \ge d_{\max}, \\
\end{array}
\right .
\label{eqn:repulsive}
\end{equation}

\noindent where
\[dt=
\left\{
\begin{array}{ll}
\textrm{min}\{t-p^\uparrow_{ij}, n^\downarrow_{ij}-t\} & \textrm{when connected}, \\
0 & \textrm{when not connected}. \\
\end{array}
\right .
\]
Furthermore, $G$ is an intensity constant, $\epsilon_0$ is small strictly positive constant to keep this force bounded, $d_{\max}$ is a cutoff distance beyond which this force no longer acts, and $\alpha$ is a parameter that determines how this force's intensity decreases with distance.

When $i$ and $j$ are not in contact, this force acts as a classical Coulomb force pushing nodes apart (left part of constraint~(\ref{eqn:not_contact_cnstr})). When nodes $i$ and $j$ come into contact, $dt$ is equal to $0$; as the contact lasts, $dt$ first increases, thereby diminishing the repulsive force, before decreasing to $0$ as $t$ approaches $n^\downarrow_{ij}$. Nodes $i$ and $j$ will therefore be drawn closer during the first part of the contact before being progressively pushed apart soon before the contact ends. This corresponds to the left part of constraint~(\ref{eqn:contact_cnstr}).

\subsubsection{Drag force}
In some cases, particularly on flat surfaces with no obvious borders or constraints (e.g., the Infocom contact trace in Section~\ref{subsec:infocom}), it may be desirable to add a drag force that will reduce to $0$ the speed of isolated nodes, thus preventing them from moving excessively far away. The drag force has the following form:

\begin{equation}
\mathbf{F}^{d}_{i} = -D \mathbf{v}_i,
\label{eqn:drag}
\end{equation}

\noindent where $D$ determines how strong the drag is and $\mathbf{v}_i$ is the current speed of node $i$.

\subsection{Issues and usage}
\label{subsec:usage}

\paragraph{Maximum speed}
While the attractive force (Eq.~(\ref{eqn:attractive})) prevents nodes from moving apart from each other too quickly, it alone does not enforce the maximum speed constraint (Constraint~(\ref{eqn:spd_cnstr})). However, this constraint can be enforced while integrating the movement equations, which we do in the rest of this paper. Interestingly, this also seems to smooth the overall inferred movement and improve the heuristic's accuracy.

\paragraph{Relaxed constraints}
When animating the graph, one must keep in mind that the goal is for each node to be within transmission range $r$ of its current contacts, and outside of range of all the other nodes. This should be encouraged but not strictly enforced, as it may otherwise lead to impossible configurations for real-life contact traces. Allowing the possibility of inexactitudes (i.e., adding or removing links in the inferred contact trace) is the price to pay for being able to infer movement from real-life traces.

\paragraph{Parameters}
The forces presented in the previous section have a number of parameters. Some of their values are inevitably somewhat arbitrary but this paragraph explains how they were chosen. A cluster of nodes can collectively have a strong repulsive force. As such, if another node comes into contact with one node in the cluster, they may never come into transmission range of each other, despite the attractive force. Setting a strong rigidity constant and setting the equilibrium length of the spring force to a point within the transmission range shown in Eq.~\ref{eqn:attractive} offsets this. In the rest of this paper, we use $K=100$. In Eq.~\ref{eqn:repulsive}, we set $\alpha=\frac{3}{2}$ and $\epsilon_0=1$. This ensures that the repulsive force (i) is bounded by $G$ and (ii) does not decrease too quickly. Finally, we choose $G$ in Eq.~\ref{eqn:repulsive} so that the equilibrium distance between two nodes, absent all other forces, is $\frac{3}{4} r$ at the exact moment they come into contact.

\paragraph{Disconnected components} An issue not usually addressed in the graph drawing community is how to deal with disconnected components. Since we are handling DTN contact traces, we cannot avoid this problem, as the connectivity graph is almost always split into several disconnected components and many isolated nodes. Freivalds et al. propose laying out each connected component separately and then using a \textit{packing} algorithm to place them relatively to one another~\cite{graph_packing}. However, this completely ignores that, in our case, the relative placement of connected components should not be arbitrary. Fortunately, our forces circumvents this problem by creating attractive forces between disconnected components and thus guiding their relative movement, orientation, splits, and merges.

\paragraph{Influence of the future} The value of $\tau$ in Eq.~(\ref{eqn:attractive}) is a trade-off. Small values of $\tau$ mean that only the very short term future is considered for animating the contact trace, while large values can create so many constraints that no movement is possible. Good values are linked to the characteristic evolution time of the connectivity graph. Finally, cutting off the attractive force eliminates long range interactions between nodes that could interfere with the initially weak attractive forces. In the rest of this paper, we use $d_{max}=2r$.

%% file: evaluation_synthetic.tex
\section{Evaluation 1: Synthetic traces}
\label{sec:evaluation_synthetic}

In this section, the heuristic described above is applied on a synthetic mobility model where the contact trace is considered perfect. Note that our heuristic makes absolutely no assumptions about the underlying mobility model. In fact, it is particularly poorly adapted to Random Waypoint (RWP), which we have chosen as the reference model, since in it the nodes try to avoid each other, whereas in RWP nodes pay no attention to each other. Nevertheless, we still manage to infer \textit{plausible} movements.

The implementation was done in Java~1.6 and run on a 2.66~GHz Intel Core~2 Duo. Videos and animations comparing original and inferred mobility, for both synthetic and experimental contact traces, are available online.\footnote{\url{http://www-npa.lip6.fr/~whitbeck/plausible.html}}

\subsection{Overview}
\label{subsec:sync_setup}

The synthetic mobility scenario considered in this section consists of 50 nodes moving according to the RWP model on a 1,000m $\times$ 1,000m torus, so as to avoid any border effects. Generic RWP has some well known shortcomings, such as a usually non-uniform steady state spatial distribution of nodes and gradual speed decay~\cite{rwp_harmful}. However, in this work, we use ``perfect simulation'' of the RWP by sampling the initial simulation state from the stationary regime~\cite{leboudec05}. In particular, we use the methods first proposed by Navidi et al. to sample initial speeds, positions, and progress along paths from the steady-state distribution~\cite{navidi04}. Note that the steady-state spatial distribution of nodes moving according to RWP on a torus is in fact uniform. Each run lasts 1,000 seconds, but this section's results are usually truncated at 900 seconds. We do this to emphasize our heuristic's strengths. Indeed, its accuracy derives essentially from its knowledge of the future, which gets progressively worse as it approaches the arbitrary 1,000-second end of the RWP simulation. Other synthetic mobility models are considered in Section~\ref{subsec:other_models}.

\begin{figure}[t]
   \centering
   \includegraphics{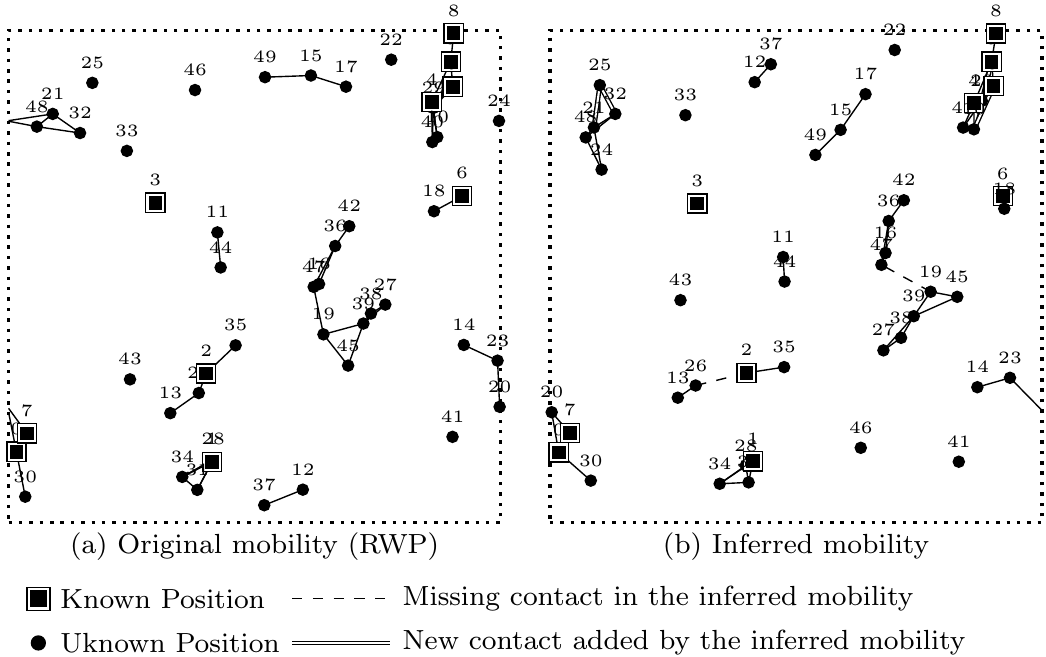}
   \caption{Mobility snapshots of the original Random Waypoint movement (a) and the movement inferred from its contact trace (b) after 450 seconds. Nodes 0 through 9 have known positions. The other 45 nodes move on a torus.}
   \label{fig:example}
\end{figure}

Unless otherwise noted, nodes move at speeds chosen uniformly between 1m/s and 10m/s (i.e., $v_{\max}=10$m/s) with no pause time, and their transmission range, $r$, is 100m. For each run of this mobility scenario, a contact trace was extracted from successive snapshots of the nodes' positions with a certain sampling period, by simply considering that any pair of nodes within transmission range of one another is in contact. A snapshot of this mobility is shown in Fig.~\ref{fig:example}a. More sophisticated radio models were not considered at this stage because our goal, in this section, is to evaluate the heuristic's behavior and the influence of its parameters when constraints~(\ref{eqn:contact_cnstr}) and~(\ref{eqn:not_contact_cnstr}) are known with perfect certainty. Of course, in real-life situations such as such of Section~\ref{sec:evaluation_reallife}, radio or MAC-layer issues lead to imperfect knowledge of these constraints.

Through some experimentation, the $\tau$ parameter of the anticipated attraction force is set to $50$. Since $v_{\max}$ is equal to $10$, this means that future contacts start significantly pushing their nodes closer to each other $5$ seconds before the contact actually appears (see Eq.~(\ref{eqn:contact_cnstr})). Smaller values meant that contacts scheduled to appear in the original trace would only show up later in the inferred trace, because of the delay until both nodes would get into transmission range. Greater values meant that each node would be attracted to a larger subset of the other nodes. If too big, this can result in preventing most movement.

Furthermore, we assume that the positions of a subset of the nodes are known at all times. We will refer to these as \textit{reference} nodes. These could be special GPS-equipped nodes, wireless access points, or 3G base stations, and help place the other nodes relatively to them. Additionally, the initial positions of all nodes were known to our heuristic. This is a very strong assumption, which allows us to skip the transitory time during which the accuracy gradually improves. A detailed analysis of this transitory time in the absence of any initial knowledge can be found in Section~\ref{subsec:sync_no_init}.

All results in this section are averaged over 20 runs. 90\% confidence intervals where systematically calculated and, clarity permitting, shown in this section's results. Our heuristic is very fast and each run typically completed in 2 to 4 minutes. One intuitive idea is common to all the results in this section: \textit{more information, better accuracy}. Indeed, since the mobility inference problem is so under-defined, any increase in information, e.g., through the presence of reference nodes or a more intense contact process, improves the accuracy of the inferred mobility.

\subsection{Number of reference nodes}
\label{subsec:sync_num_nodes}

\begin{figure}[t]
  \centering
  \subfloat[0 known nodes (0.57)\label{subfig:corr_matrix_kn0}]{\scalebox{0.9}{\includegraphics{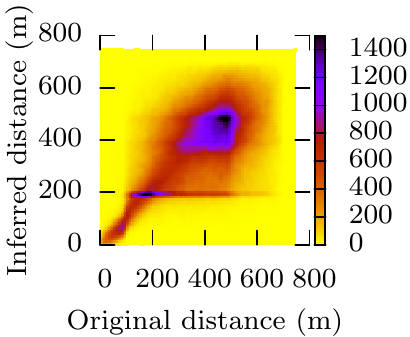}}} \quad
  \subfloat[5 known nodes (0.84)\label{subfig:corr_matrix_kn5}]{\scalebox{0.9}{\includegraphics{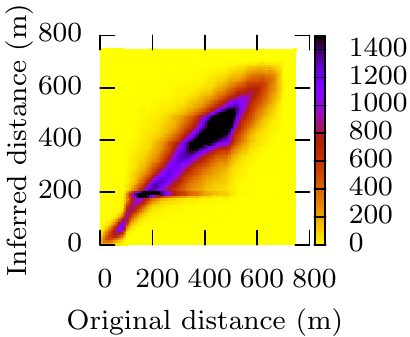}}} \quad
  \subfloat[25 known nodes (0.93)\label{subfig:corr_matrix_kn25}]{\scalebox{0.9}{\includegraphics{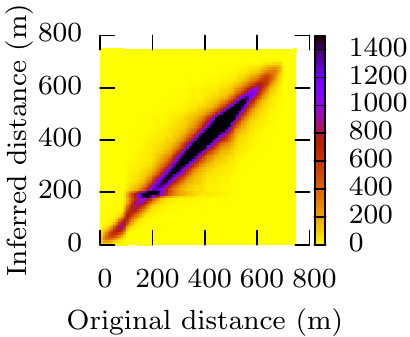}}}
  \caption{Correlation between pairwise distances in a synthetic mobility trace and its inferred mobility. Correlation values are given in parentheses.}
  \label{fig:pairwise_dist_corr}
\end{figure}

\begin{figure}[t]
	\centering
	\subfloat[Pairwise Correlation\label{subfig:num_known_corr}]{\includegraphics{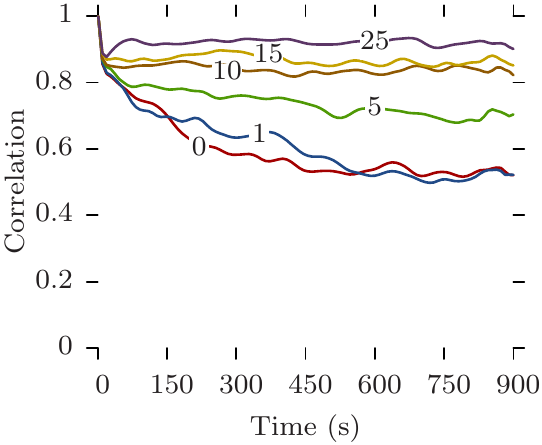}} \quad
	\subfloat[Mean Distance Error\label{subfig:num_known_mde}]{\includegraphics{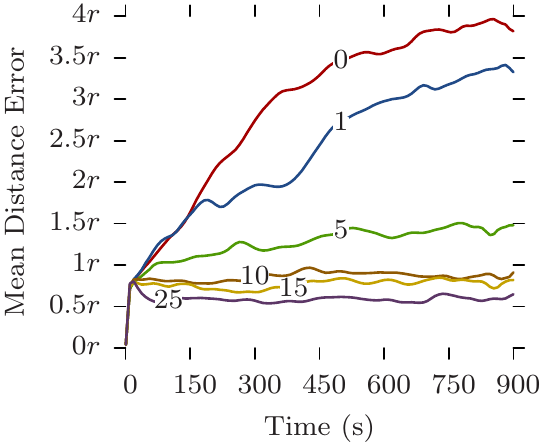}}
	\caption{Influence of the number of known node positions on accuracy.}
	\label{fig:num_known}
\end{figure}

In this section, we run the above scenario for different numbers of reference nodes. We first compare the inferred mobility to the original. Generally, while we do not expect to be able to infer the \textit{exact} node positions, the \textit{relative} distances between pairs of nodes should be correlated. Indeed, Fig.~\ref{fig:pairwise_dist_corr} examines the correlation between the pairwise distances in the original and inferred mobility. Every time step, the distance between each pair of nodes was measured in both the original and inferred mobility. Fig.~\ref{subfig:corr_matrix_kn0} represents the correlation scatter plot over the entire simulation in the absence of any reference nodes. Figs.~\ref{subfig:corr_matrix_kn5} and~\ref{subfig:corr_matrix_kn25} are obtained when the positions of 5 and 25 reference nodes are known, respectively. Naturally, only pairs of nodes in which both were not reference nodes were considered. The progression in Fig.~\ref{fig:num_known} demonstrates how increasing the number of reference nodes improves the pairwise distance correlation between original and inferred mobility, although with diminishing returns.

One artifact of our heuristic is visible in Fig.~\ref{fig:pairwise_dist_corr}. Indeed, a horizontal ``bar'' appears at 200m in the inferred mobility. This is particularly visible on Figs.~\ref{subfig:corr_matrix_kn0} and~\ref{subfig:corr_matrix_kn5}. This means that a number of distances above 200m in the original mobility are roughly mapped to precisely 200m in the inferred mobility. This is a result of the cutoff distance of the repulsive force as discussed in Section~\ref{subsec:usage}.

Fig.~\ref{subfig:num_known_corr} plots the pairwise correlation over time instead of averaged over the entire movement, for different numbers of reference nodes. All curves begin at 1 but there is a clear difference between the situation with 0 reference nodes, where the pairwise correlation steadily grows worse, and the situation with 25 reference nodes, where the pairwise correlation is steady around 0.9. Much of this ``stabilization'' has already occurred with about 10 reference nodes (i.e., 20\% of the total number of nodes).

Not only does \textit{relative} pairwise distance correlation improve with the number of known nodes, but \textit{absolute} positioning accuracy does so as well. This is visible on Fig.~\ref{subfig:num_known_mde}, which plots the mean distance error over time. If $N$ is the set of all nodes, $R$ the set of reference nodes, and $\mathbf{r}_i(t)$ (resp. $\hat{\mathbf{r}}_i(t)$) the position of node $i$ in the original (resp. inferred) mobility at time $t$, then the mean distance error at time $t$ is $\frac{1}{|N \setminus R|} \sum_{i \in N \setminus R} \| \hat{\mathbf{r}}_i(t) - \mathbf{r}_i(t) \|$. In Fig.~\ref{subfig:num_known_mde}, the mean distance error increases steadily to $3.8r$ when no nodes have known positions. Note that the expected mean distance error for a randomly placed node on a 1,000m by 1,000m torus is roughly 382m, i.e., $3.8r$.\footnote{The expected mean distance error for a node uniformly randomly placed on a $h \times w$ torus, is $\frac{1}{hw}\int^{h/2}_{-h/2} \int^{w/2}_{-w/2} \sqrt{x^2+y^2} dx dy$.}. Unlike the correlation scores on Fig.~\ref{subfig:num_known_corr}, the curves for $0$ and $1$ reference nodes are distinct. This suggests that even though relative node positioning may be declining in both cases, knowing the position of at least one nodes helps in inferring the absolute node positions of all the others. When no such information is available, the absolute positioning accuracy of the inferred mobility degrades very quickly.

The mean distance error converges below $r$ with $10$ reference nodes and approaches $0.5r$ with $25$ such nodes. It is difficult for our heuristic, based solely on contact information, to achieve precisions greater than $O(r)$. Indeed, in the sparse scenario considered, nodes are rarely even connected to a single reference node. This is visible, for $10$ reference nodes, on Fig.~\ref{fig:example}a. Most of the positioning is therefore achieved through multi-hop inference using past, present, and future contacts. In fact, given how little information we use, the achieved absolute positioning accuracy is surprisingly good.

\subsection{Node density}
\label{subsec:sync_density}

\begin{figure}[t]
	\centering
	\subfloat[Pairwise Correlation\label{subfig:num_nodes_corr}]{\includegraphics{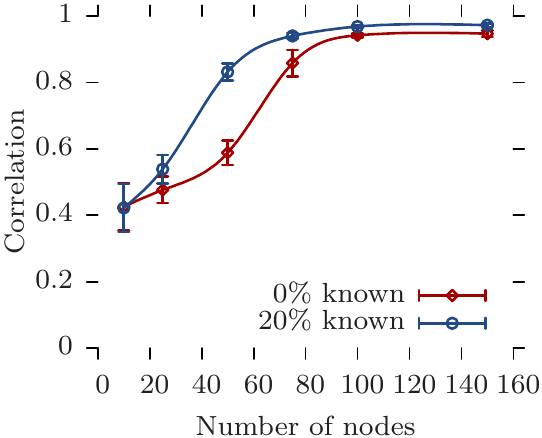}} \quad
	\subfloat[Mean Distance Error\label{subfig:num_nodes_mde}]{\includegraphics{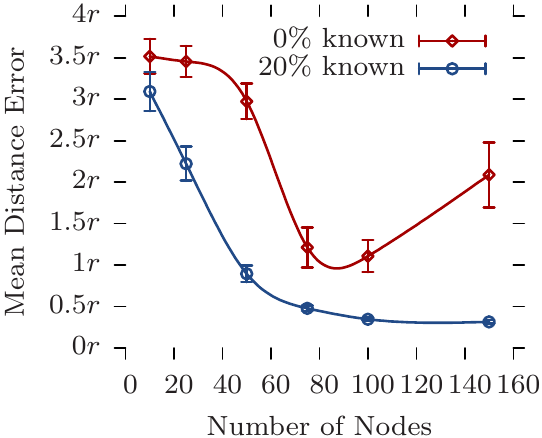}}
	\caption{Influence of the network density on accuracy. Known percentage refers to the fraction of nodes whose positions are known.}
	\label{fig:num_nodes}
\end{figure}

Knowledge of past and future contacts is key to accurately positioning nodes. Therefore we should expect denser scenarios with a higher contact intensity to provide more information and thus improve our heuristic's accuracy. Indeed, in an extremely sparse scenario, nodes may remain completely isolated for a long period of time. The exponential term in attractive force (Eq.~(\ref{eqn:attractive})), will remain close to $0$ most of the time, thus letting the nodes merely drift relatively to one another. Conversely, we expect a very dense scenario to provide very fine-grained information on past, present, and upcoming contacts which should greatly constrain possible movement and improve the overall accuracy.

In order to test this idea, we modify the previous scenario by simply modifying the total number of nodes. For each total number of nodes, we run two scenarios: one with no reference nodes and another with 20\%.

The results are displayed on Fig.~\ref{fig:num_nodes}. Fig.~\ref{subfig:num_nodes_corr} plots the average pairwise distance correlation over the entire movement while Fig.~\ref{subfig:num_nodes_mde} does the same for the mean distance error. With 20\% of reference nodes, these results confirm our intuition that greater node density leads to more accurate mobility inference. In extremely sparse scenarios with only $10$ nodes, the accuracy ($3r$) is not much better than random placement (roughly $4r$, see Section~\ref{subsec:sync_num_nodes}). However, in denser, completely connected scenarios with close to $100$ nodes, the accuracy is close to $0.25r$.

Interestingly, in the absence of any reference nodes, two separate trends are at work. On the one hand, greater node density still leads to improved pairwise distance correlation with tight confidence intervals (Fig.~\ref{subfig:num_nodes_corr}). On the other hand, the mean distance error is very unstable, leading to larger confidence intervals and an error \textit{increase} for $150$ nodes (Fig.~\ref{subfig:num_nodes_mde}). Recall that with no reference nodes, the mean distance error increases over time until reaching roughly $4r$ (see Section~\ref{subsec:sync_num_nodes}). Increasing the node density merely changes the \textit{speed} at which this happens. More nodes mean more constraints and, given correct initial positions, this prevents nodes from collectively drifting away from their correction positions too quickly. This accounts for the initial improvement in mean distance error. However, increased node density eventually leads to a fully connected network, where \textit{local errors} quickly ripple through the \textit{entire topology}. Therefore an early error can create a jump in mean distance error as it drags all the other nodes with it. This accounts for the widening confidence intervals and the increasing absolute positioning error with more than $100$ nodes, in the absence of reference nodes.

\subsection{Contact trace randomization}
\label{subsec:sync_rand}

\begin{figure}[t]
	\centering
	\subfloat[Average percentage of added contacts\label{subfig:sync_async_added}]{\includegraphics{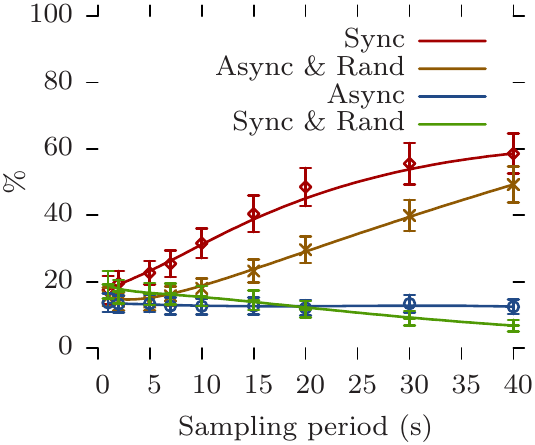}} \quad
	\subfloat[Average percentage of missed contacts\label{subfig:sync_async_removed}]{\includegraphics{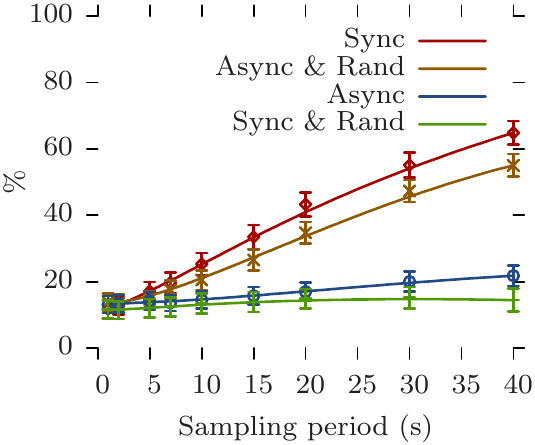}}
	\caption{Effects of contact trace randomization on inference accuracy. At any given time, added and missed contacts are expressed as a percentage of the number of existing contacts at that time.}
	\label{fig:sync_async}
\end{figure}

In this section, we examine various ways of extracting and handling contact traces. Whereas up until now, we have been comparing the original and inferred \textit{mobility} directly, in this section we compare original and inferred \textit{contacts}. Every time step, we check the constraints~(\ref{eqn:contact_cnstr}) and~(\ref{eqn:not_contact_cnstr}). If two nodes that should be in contact are not within transmission range of each other, that contact is considered \textit{missed}. Conversely if two nodes not in contact are within transmission range of each other, an \textit{added} contact is counted. This is a rather strict way of comparing two contact traces, as even slight delays in the start or end of a contact will register as respectively a missed or added contact for that time period.

When extracting contacts from a given movement, there are several ways to proceed. The most straightforward method is to take snapshots of the connectivity graph every $T$ seconds. We will refer to this as the \textit{synchronous} method. A more subtle method involves simulating an actual contact trace experiment in which each node periodically scans its neighborhood. Since each node has a random starting time, we will refer to this as the \textit{asynchronous} method. This method, in fact, corresponds to the way the real-life iMote experiments considered in Section~\ref{sec:evaluation_reallife} were collected.

Conversely, when using a contact trace with a known sampling period $T$, one can either elect to use it ``as-is'', or randomize the event times a little. Indeed, if events are measured with a sampling period $T$, then an event (link-up or link-down) measured at time $t$, really occurred between $t$ and $t-T$. In this paper, we randomize a contact trace with sampling period $T$ by shifting all events back by a time uniformly chosen between $0T$ and $0.8T$.

Fig.~\ref{fig:sync_async}, plots the percentage of \textit{added} and \textit{missed} contacts for different sampling periods and for different combinations of synchronous or asynchronous sampling and randomization.

In the \textit{synchronous} approach, the sampling period of the original contact trace has an important impact on the quality of our inference. The proportion of both added and missed contacts increases with the sampling period. This is due to several reasons. Firstly, as the sampling period increases, it becomes more difficult to assume that a contact present in one period but not the next lasted the whole period. Lacking other information, our heuristic does however make this assumption. During a time period, a given node, when pushed by the anticipated attraction force towards its next contact, is still restrained by the attractive forces of the nodes that were in contact with it at the beginning of the period. For longer sampling periods, we may overestimate the duration of many contacts and therefore prevent a node from moving towards its future contacts. This translates into both \textit{missed} contacts, from not getting within transmission range of new contacts in time, and \textit{added} contacts from remaining within transmission range of old contacts. Secondly, smaller sampling periods also catches short contacts that would otherwise be ignored. These provide many extra contacts that a node must meet on the way to meeting its next contacts according to the longer sampling period, and thus enable a much smoother and progressive mobility inference.

However, when contacts are measured \textit{asynchronously}, the number of contacts added or missed remains between 10\% and 20\% regardless of the sampling period. This due to two reasons. Firstly, having each node scan its neighborhood with a different phase results in a continuous, though incomplete, stream of information on future contact events, thus providing a much finer-grained picture of the contact trace than implied by the sampling period. Furthermore, each link is being monitored by a pair of nodes, and the time of an event on that link is measured as the minimum of the times at which it was noticed by each node in the pair. This further increases the precision of the measured contact trace. As most real-life contact traces are measured in this fashion, this is encouraging for the applicability of our heuristic to existing real-life traces (Section~\ref{sec:evaluation_reallife}).

Interestingly, contacts randomization has very different effects on \textit{synchronous} and \textit{asynchronous} contacts measurements. Indeed, it considerably improves the quality of the movement inferred from \textit{synchronous} sampling, making it comparable to that of the \textit{asynchronous} approach. On the other hand, when randomization is used with \textit{asynchronous} sampling, it actively degrades the quality of the inferred mobility.

\subsection{No knowledge of initial positions}
\label{subsec:sync_no_init}

\begin{figure}[t]
	\centering
	\subfloat[Pairwise Correlation\label{subfig:rand_num_known_corr}]{\includegraphics{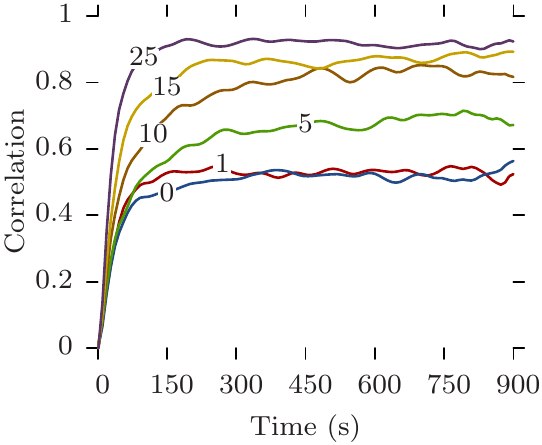}} \quad
	\subfloat[Mean Distance Error\label{subfig:rand_num_known_mde}]{\includegraphics{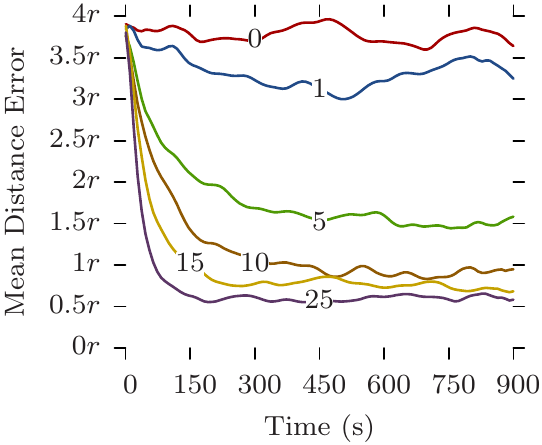}}
	\caption{Transitory regime when initial positions of nodes are unknown.}
	\label{fig:rand_num_known}
\end{figure}

Up until now, all our simulations assumed that the initial positions of all nodes were known. This conveniently allowed us to isolate the parameters we wished to test (e.g., number of nodes with known positions and node density), and ignore the effects of a transitory regime during which the inferred movement converges over time towards the actual original movement. In this section we assume that the initial positions of all nodes are random, and examine how this impacts our results. We also show that by temporarily relaxing the maximum speed constraint, convergence becomes much faster.

Figs.~\ref{subfig:rand_num_known_corr} and \ref{subfig:rand_num_known_mde} describe, respectively, the pairwise distance correlation and the mean distance error when starting from initially random positions. All simulation parameters being exactly identical to those in Section~\ref{subsec:sync_num_nodes}, they should be compared to Figs.~\ref{subfig:num_known_corr} and \ref{subfig:num_known_mde}. Recall that the expected initial mean distance error on a 1,000m by 1,000m torus is roughly 382m, i.e., $3.8r$ (see Section~\ref{subsec:sync_num_nodes}).

As expected, both the correlation and the mean distance error improve over time and converge towards the values obtained when the initial positions are known. However this convergence is much faster when more nodes have known positions. For example, with $15$ known nodes it takes about $200$ seconds to stabilize around the same pairwise correlation and mean distance error values as when the initial positions are known. The convergence is faster with $25$ nodes, whereas $10$ and $5$ are slower. Unsurprisingly, in the absence of any reference nodes, the mean distance error never improves over a random guess (Fig.~\ref{subfig:rand_num_known_mde}). However, the pairwise correlation does initially improve before stabilizing around $0.5$ (Fig.~\ref{subfig:rand_num_known_corr}). After the transitory period, the accuracy achieved in each case is the same as that obtained when the initial positions of all nodes are known (Figs.~\ref{subfig:num_known_corr} and~\ref{subfig:num_known_mde}).

These results indicate that a certain proportion of nodes with known positions (in this example, around 20\%) are needed to achieve a good and stable accuracy. When such reference nodes are present, the assumption that initial positions are known allows us to eschew the transitory regime and does not fundamentally change the results.

\begin{figure}[t]
	\centering
	\subfloat[Pairwise Correlation\label{subfig:cheat_rand_num_known_corr}]{\includegraphics{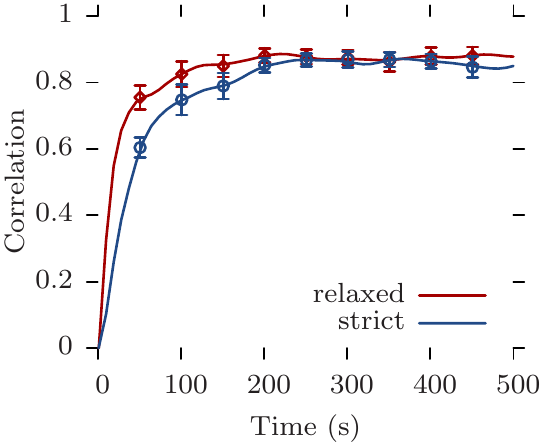}} \quad
	\subfloat[Mean Distance Error\label{subfig:cheat_rand_num_known_mde}]{\includegraphics{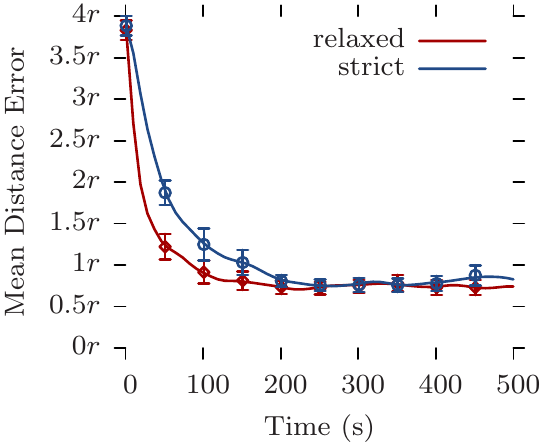}}
	\caption{Improving convergence time by relaxing the maximum speed constraint for $50$ seconds when initial positions of nodes are unknown. Here, $15$ out of $50$ nodes are reference nodes, i.e., with known positions.}
	\label{fig:cheat_rand_num_known}
\end{figure}

Furthermore, there are several ways in which the duration of the transitory regime can be shortened. For example, temporarily relaxing the maximum speed constraint enables the nodes to find their correct positions faster. Fig.~\ref{fig:cheat_rand_num_known} compares the pairwise correlation and the mean distance error with $15$ reference nodes when the maximum speed is enforced or relaxed. In the latter case, during the initial $50$ seconds, nodes were allowed to move according to the accelerations determined by the force-based heuristic without any restrictions on maximal speed. Therefore a little initial ``cheating'', can significantly shorten the transitory regime.

\subsection{Other Mobility Models}
\label{subsec:other_models}

\begin{figure}
  \centering
  \subfloat[Pairwise Correlation\label{subfig:other_corr}]{\includegraphics{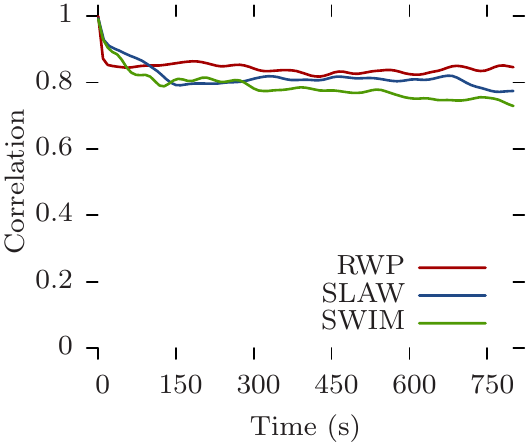}} \quad
  \subfloat[Mean Distance Error\label{subfig:other_mde}]{\includegraphics{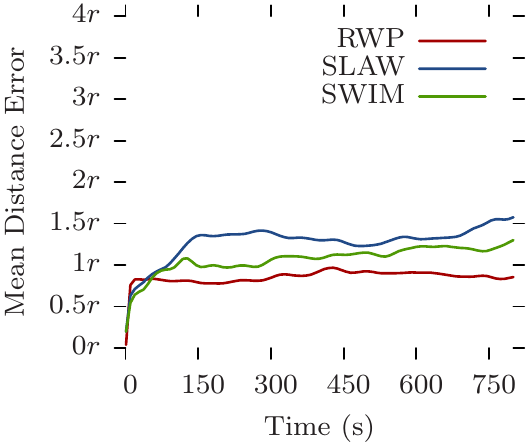}}
  \caption{Inferred mobility accuracy for different mobility models. (50 nodes including 10 with known positions)}
  \label{fig:other}
\end{figure}

While the evaluation above focused on Random Waypoint, our plausible mobility heuristic of course works with different synthetic mobility models. In this section we consider three recent state-of-the-art mobility models: (i) SLAW (Self-similar Least Action Walks~\cite{Lee2009}), (ii) SWIM (Small World in Motion~\cite{Mei2008}), and (iii) TVC (Time-Variant Community Model~\cite{Hsu2009}). SLAW generates fractal waypoints and then plans trips while trying to minimize overall travel distance. SWIM is based on the intuition that we spend long amounts of time in a few popular areas (e.g., home or office) and small amounts of time in the others, and that the further the destination the faster the travel speed. TVC captures non-homogeneous behaviors in both time and space by using preferred communities that change at successive time \textit{epochs}. Unfortunately, the TVC model exhibits a ``hopping'' behavior where nodes ``teleport'' to their new favored communities at the beginning of each new epoch. This completely throws off our heuristic, which involves gradually bringing together nodes before their next contact comes up. It is therefore not included in this section's results.

We use the default parameters for both SLAW and SWIM except for the number of nodes (50), the maximum speed (10m/s), and the simulation area (1000m $\times$ 1000m). The RWP results are those of our default scenario. In order for the results to be comparable, we did not try to adjust the parameters of our plausible mobility heuristic to each type of mobility. The exact same parameters were used for inferring mobility from RWP, SLAW, and SWIM contact traces. Fig.~\ref{fig:other} shows the accuracy over time for all three models. The results are averaged over 20 runs. As RWP, the accuracy for SLAW and SWIM quickly stabilizes, albeit at slightly worse values. This is primarily due to two factors. Firstly, both SWIM and SLAW have pause times, whereas our RWP runs did not. This results in an overall reduction of the intensity of the contact process and thus less information on the future. Secondly, SWIM and, even more so, SLAW exhibit spatial inhomogeneity. This means that certain nodes can spend long periods of time without coming close to a reference node.

%% file: evaluation_reallife.tex
\section{Evaluation 2: Real-life traces}
\label{sec:evaluation_reallife}

Ideally, evaluation should be conducted against traces that measure both contacts and node locations but, unfortunately, no such traces are available. We also chose not to use existing GPS traces because, as discussed in Section~\ref{sec:related}, these require significant pre-processing (e.g., scaling, adding new nodes, time-shifting) before they can produce enough contacts to make our heuristic relevant. For example, let us consider the UMass Dieselnet GPS traces, composed of 30-40 buses in an area covering approximately 150 square miles~\cite{UMassDieselNet}. If we were to use a sufficiently long transmission range (e.g., $~$1.5 km)  for each bus to be, on average, within range of one other, we estimate, based on the results in the previous section, that our localization accuracy would be around 3km. Extremely sparse GPS traces therefore lead to unusable localization information.

Therefore, in this section, we instead focus on real-life \emph{contact traces}, specifically  the Infocom 2005 dataset~\cite{chaintreau_mobility} and the Rollernet dataset~\cite{tournoux08_rollernet}. The Rollernet trace is particularly interesting because of its short sampling period (see Table~\ref{table:traces}) and relatively constrained environment.

In these cases, the ground truth (i.e., the original node mobility) is obviously not available, hence we cannot evaluate how accurate the inferred mobility is. However it is still possible to compare original and inferred \textit{contacts}. Compared to the synthetic mobility in the previous section, such real-life contact traces results must be considered lossy, i.e., the absence of a contact in the dataset does not imply that two nodes weren't within transmission range of each other. Such lossy information may even lead to an \textit{impossible} system of constraints in the strict sense of Section~\ref{sec:formal}. In this case, our force-based heuristic will tend to place some nodes close to each other even though there is no registered contact in the trace, thereby adding many new contacts. Furthermore, their sampling periods ($120$ and $15$ seconds, respectively) are significantly longer than the $1$ second sampling period previously used in the synthetic scenarios. However, since the sampling periods are asynchronous (see Section~\ref{subsec:sync_rand}), we expect this to have a smaller impact on accuracy than the lossy measurements.

\subsection{Infocom}
\label{subsec:infocom}

\begin{figure}
	\centering
	\subfloat[Inter-contact times\label{subfig:infocom_ict}]{\includegraphics{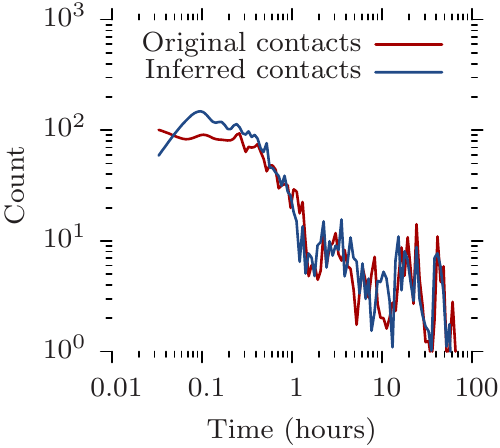}} \quad
	\subfloat[Inter-contact time distribution (CDF)\label{subfig:infocom_ict_cdf}]{\includegraphics{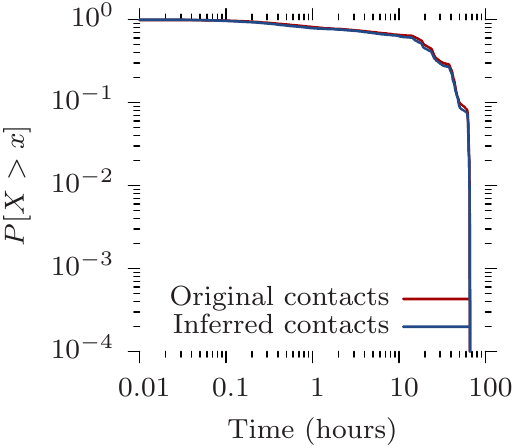}}
	\caption{Comparison of original and inferred contacts in the Infocom 2005 dataset~\cite{chaintreau_mobility}. Values were smoothed over 120-seconds time periods.}
   	\label{infocom_ict}
\end{figure}

The Infocom dataset used Bluetooth Intel iMotes to capture the contacts between participants at the Infocom 2005 conference. Our plausible mobility algorithm ran on the entire contact trace using the same parameters as previously, except for $\tau$. Previously, with a $1$ second sampling period, $\tau$ was equal to $50$ which meant that upcoming events started having a significant influence on movement $5$ seconds before they actually occur. With a the Infocom dataset's $120$ second sampling period, we increased $\tau$ to $500$. The nodes were placed on a boundless planar area. The anticipated attraction forces generally prevent nodes from moving too far away. However, this was not sufficient and a drag force (Eq.~(\ref{eqn:drag})) was added to gradually reduce the speed of isolated nodes to zero.

Fig.~\ref{subfig:infocom_ict} plots the raw inter-contact distribution of both the original and inferred contact traces. It shows that the inferred mobility tends to increase the number of short inter-contacts but then closely follows the original contact trace for longer inter-contact times. This is not the case for the Rollernet inferred mobility, discussed below. On average, as a percentage of the total number of contacts at each given time, the inferred mobility adds 16.3\% extra contacts, many of them short-lived.

Interestingly, these extra short inter-contact time are comparatively few compared to the longer inter-contact times and thus have nearly no impact on the inter-contact time cumulative distribution function. Indeed, in Fig.~\ref{subfig:infocom_ict_cdf}, the original and inferred contacts cumulative distribution function are nearly identical.

These results show that the proposed mobility is clearly \textit{plausible}, and could have produced the original contact trace, or at least one very similar. What about the added short inter-contact times?  Are we introducing bogus contact opportunities? Or are we identifying contacts that the experiment missed? Or both? Bluetooth contact traces have long sampling periods and an intrinsic lack of reliability, thus making it difficult to decide. However, the results in the following section, based on the Rollernet contact trace in which node mobility has many more constraints, lead us to believe that the we may be detecting many contact opportunities missed by the Bluetooth measurements.

\subsection{Rollernet}
\label{subsec:rollernet}

\begin{figure}
	\centering
	\subfloat[Inter-contact times\label{subfig:rollernet_ict}]{\includegraphics{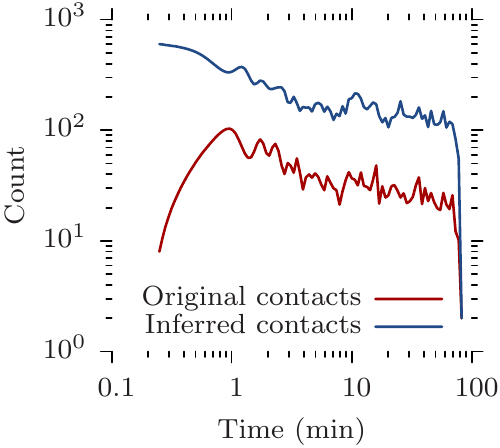}} \quad
	\subfloat[Inter-contact time distribution (CDF)\label{subfig:rollernet_ict_cdf}]{\includegraphics{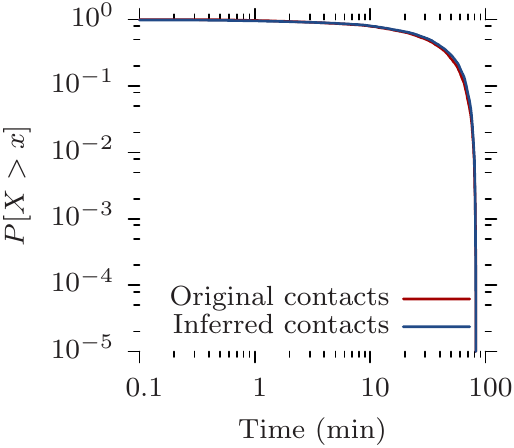}}
	\caption{Comparison of original and inferred contacts in the Rollernet dataset~\cite{tournoux08_rollernet}. Values were smoothed over 15-seconds time periods.}
   	\label{fig:rollernet_ict}
\end{figure}

The Rollernet contact trace~\cite{tournoux08_rollernet} was collected using Bluetooth Intel iMotes during a rollerblading tour around Paris. It has the shortest sampling period, $15$ seconds, of all Bluetooth contact traces, and therefore comes closest to capturing the evolution of the connectivity graph. Furthermore, the nodes are highly mobile and the average contact duration (26 seconds) is barely longer than the sampling period (15 seconds).

When inferring mobility for the Rollernet contact trace, we used the same parameters as in the synthetic simulation with the following modifications. There are no fixed anchor nodes in the trace, but the head and tail nodes of the rollerblading tour are known. No participant skated ahead (resp. behind) the head (resp. tail) node. The head and tail nodes were constrained to moving along a horizontal axis. If ever a node wants to pass the head node for example, then the head node is moved to ensure that it remains ahead. This enables all nodes to be placed relatively to the head and tail nodes. Furthermore, since the rollerbladers go down Paris streets, we can approximate the shape of the crowd as a 50-meter wide rectangle.\footnote{The largest avenue on the path of the rollerblading tour is the 70-meter wide (including sidewalks) Champs-Elys\'ees.} This width constraint is added to our heuristic in the form of a potential well, i.e., walls at $\pm 25$ meters that push nodes away. Finally, the iMotes have a relatively short transmission range ($30$ meters) and the heuristic's range parameter is set accordingly. Our plausible mobility heuristic ran on the first $5,000$ seconds of the contact trace.

Due to the rapid contact process in the Rollernet trace, the anticipated attraction forces are sufficient to keep the rollerblading tour compact and naturally lead to the emergence of the accordion phenomenon~\cite{tournoux08_rollernet} because the head and tail node get closer when the contact density increases and pushed apart when it decreases. The resulting mobility is aesthetically pleasing and helps guide intuition when working on the dataset.

Fig.~\ref{subfig:rollernet_ict} plots the raw inter-contact distribution of both the original and inferred contact traces. Not only does the inferred mobility add a huge number of short inter-contacts, but, unlike in the Infocom example, also adds many longer inter inter-contacts as well. On average, the inferred mobility more than doubles the number of existing contacts at any given time (116.3\% of extra contacts as a percentage of the total number of contacts at each given time). Several things may account for this. Firstly, the Infocom mobility, consisting of participants in a conference, attending sessions, meeting for lunch and buffets, is intrinsically a lot more static than the very dynamic mobility of participants in a rollerblading tour. Therefore, the Rollernet contact trace is probably ``lossier'' than the Infocom one. Secondly, the 50-meter width constraint imposed in the Rollernet inferred mobility, practically forces any two rollerbladers passing one another to create a contact opportunity. In the Infocom example above we placed no constraints on the spatial extension of the inferred mobility. Indeed during the day, participants may all be in the same room, whereas later in the evening, they may be spread out around the entire city.

However, as with the Infocom trace, this does not have any influence on the inter-contact time cumulative distribution function as seen in Fig.~\ref{subfig:rollernet_ict_cdf}. This suggests that even though many new contacts are indeed added, these are consistent with the measured ones. As nodes move back and forth in the rollerblading tour, they meet other nodes and create contact opportunities that are not in the original contact trace but could plausibly have been captured with a finer measurement technique. Of course this interpretation is tentative and needs to be confirmed using more precise and reliable contact measurements, but if true, means that the \textit{plausible} mobility proposed by our heuristic is a pretty close to the original mobility, even with a lossy contact trace.

%% file: conclusion.tex
\section{Conclusion and further work}
\label{sec:conclusion}

In this paper, we examined a new and interesting problem, the inference of a \textit{plausible} mobility from a wireless contact trace. Indeed, mobility is more difficult to measure but enables movement visualization, contact-trace transformations (e.g., adding new nodes or increasing the overall density), and better simulations, particularly in dense networks. On the other hand, contact information is easier to measure but makes direct transformations of the contact trace difficult and only allows for simplistic simulation models. Our heuristic solution, based on ideas from dynamic graph drawing, can animate any wireless contact trace, while making practically no assumptions on the mobility that produced the contact trace. Our results highlight the need for reliable high-frequency contact traces in order to extract a \textit{plausible} mobility from a wireless contact trace. Furthermore, the collection of combined contacts and GPS coordinates traces, in dense environments, would provide a firm ground truth and greatly enchance our ability to evaluate mobility inferrence  techniques.

This work, a first approach of a difficult problem, can be pursued in several directions. Firstly, the heuristic presented in this paper still makes a small number of errors (i.e., added and missed contacts) on synthetic scenarios. Perhaps the plausible mobility obtained from the techniques in this paper could be used, not as the definitive solution, but as an initial ``good bet'', or first-pass, in a more complete algorithm that actually solves the large constraint system of Section~\ref{sec:formal}. Secondly, being able to infer mobility from a contact trace opens up many possibilities for transforming these, and then re-using them in scenarios other than the exact circumstances in which they were obtained.  Finally, the idea that inferring mobility from a contact trace before running simulations on it ought to lead to more realistic results than just using the contact trace alone, must be properly tested. For example, one could compare the performance of various network protocols in an opportunistic network based on simulations, with complete knowledge of the mobility, with only contact information, and with the \textit{plausible} mobility inferred from the contacts. The results on the inferred mobility should be closer to those using the real mobility than to those using only the contact information.